\documentstyle[amssymb,aps,preprint]{revtex}
\begin{document}
\draft
\title{Observation of an unusual field dependent slow magnetic relaxation and two
distinct transitions in a family of new complexes}
\author{Song Gao$^{1,*}$, Gang Su$^{2,\dagger}$, Tao Yi$^1$, Bao-Qing Ma$^1$}
\address{$^{1}$State Key Laboratory of Rare Earth Materials Chemistry and\\
Applications, PKU-HKU Joint Laboratory on Rare Earth Materials and\\
Bioinorganic Chemistry, Peking University, Beijing 100871, China\\
$^{2}$Department of Physics, Graduate School, Chinese Academy of Sciences, \\
P.O. Box 3908, Beijing 100039, China}
\maketitle

\begin{abstract}
An unusual field dependent slow magnetic relaxation and two distinct
transitions were observed in a family of new rare earth-transition metal
complexes, [Ln (bipy) (H$_{2}$O)$_{4}$ M(CN)$_{6}$] $\cdot $1.5 (bipy) $%
\cdot $ 4H$_{2}$O (bipy = 2,2'-bipyridine; Ln = Gd$^{3+}$,Y$^{3+}$; M = Fe$%
^{3+}$, Co$^{3+}$). The novel magnetic relaxation, which is quite different
from those in normal spin glasses and superparamagnets but very resembles
qualitatively those in single-molecule magnet Mn$_{12}$-Ac even if they
possess different structures, might be attributed to the presence of
frustration that is incrementally unveiled by the external magnetic field.
The two distinct transitions in [GdFe] were presumed from DC and AC
susceptibility as well as heat capacity measurements.
\end{abstract}

\pacs{PACS numbers: 75.50.Xx, 75.40.-s, 81.05.Zx}

Recently there has been growing interest in studying frustrated systems, and
a variety of magnetic systems with geometrically frustrated structures, such
as the three-dimensional (3D) cubic pyrochlore lattice [1-3], the
two-dimensional (2D) kagome lattice [4, 5], and some disordered molecular
systems [6-10], etc., have been extensively investigated. A number of new
phenomena, like noncollinear Ne\`{e}l long-range order, quantum disorder,
order by disorder, order with disorder, etc., have been predicted or
observed at low temperatures[1-10]. Those systems are not genuine spin
glasses, but they exhibit spin glass-like behaviors, showing the salient
role of frustration which is normally generated by the competition between
different kinds of interactions or by special lattice structures. Here we
report novel, unusual but universal properties in a class of new complexes
which are geometrically frustrated rare earth-transition metal cyanides.

The compounds [Ln (bipy) (H$_{2}$O)$_{4}$ M(CN)$_{6}$] $\cdot $1.5 (bipy) $%
\cdot $ 4H$_{2}$O (Ln = Gd$^{3+}$,Y$^{3+}$; M = Fe$^{3+}$, Co$^{3+}$)
(abbreviated as [GdFe], [YFe] and [GdCo], respectively) were prepared by
mixing aqueous solution of K$_{3}$[M(CN)$_{6}$] (1mmol) and alcohol solution
of 2,2'-bipyridine in a 1/3 molar ratio, then by dropping Ln(NO$_{3}$)$_{3}$
(1mmol) aqueous solution slowly without stirring. Single crystals were
obtained by slow evaporation of the solution at room temperature. Single
crystal X-ray analysis revealed that they are isomorphous, and only the
structure of [GdFe] was exemplified here. The transition metal ion Fe$^{3+}$
and lanthanide ion Gd$^{3+}$ are bridged by cyano group CN, forming a chain
as shown in Fig.1(a). Within the chain, Gd-Fe, Fe-Fe (or Gd-Gd) separations
are $5.5$ and $10.89$\AA , respectively. In {\it ab} plane, the chains are
connected through hydrogen bonds with nearest inter-chain Gd-Fe, Fe-Fe (or
Gd-Gd) separations $7.7$, $9.75$ $(=a)$ and $10.66$ $(=b)$\AA . Within the
layer, the connection of Fe$^{3+}$ ions (solid line) or Gd$^{3+}$ ions (dash
line) gives rise to a slightly deformed triangular lattice, respectively.
The nearest separation of Fe or Gd ions between the two neighbor layers are $%
9.906$, $10.81$\AA\ for A-B, and $11.44$ \AA\ for B-A, as shown in Fig.
1(b). Since the separation between the Fe$^{3+}$ (or Gd$^{3+}$) ions in the
chain ($10.89$\AA ) is comparable with the nearest inter-chain and
inter-layer separations ($9.75-11.44$\AA ), [GdFe] cannot be regarded as a
good one-dimensional compound. Furthermore, the inter-layer interactions
should be weaker than those in {\it ab} plane due to the weaker $\pi $-$\pi $
stack of pyridine rings between layers. As a result, [GdFe] is actually an
anisotropic 3D magnetic system in which the interactions along the {\it %
c-axis} are smaller than that within {\it ab} plane. It is this weak residue
interactions along the {\it c-axis} to make the system exhibit magnetic
long-range order. The physical properties of these systems are predominantly
determined by magnetic ions within the plane. Apparently, if the coupling
between Fe$^{3+}$ (or Gd$^{3+})$ ions in {\it ab} plane are
antiferromagnetic (AFM), the frustration will occur. If really so, [LnM] can
be in principle viewed as geometrically frustrated systems. Although spin
glass-like behaviors are expected for such systems, rather weak interactions
between Fe$^{3+}$ (or Gd$^{3+})$ ions (due to long separations) and the
intervention of the rare earth ions will complicate the situation.

$\chi _{M}T$ (or equivalently, the effective magnetic moment $\mu _{eff}$ $%
=(8\chi _{M}T)^{1/2}$) of [GdFe] was measured in $100$Oe to $5$kOe DC fields
with a SQUID magnetometer (Quantum Design MPMS5) with crystalline sample
confined in parafilm. The results show an unusual inverse field-dependence
of $\chi _{M}T$ below $4$K, namely, with increasing temperature it first
increases to reach a maximum, and then decreases, after undergoing a
plateau, to approach a constant value (atomic limit), as shown in Fig. 2(a).
The drops of $\chi _{M}T$ below $2.5$K, signal the existence of AFM
interactions among the metal ions. When the magnetic field is stronger than $%
5$kOe, the peak disappears and the AFM behavior is dominant. We note that
the seemingly similar behaviors of $\chi _{M}T$ were also reported in the
highly frustrated triangles-in-triangles crystalline system [4,5]. Apart
from the strong peak around $2.5$K, there is a shoulder occurring around $3$%
K especially in low fields, suggesting that there must be a phase transition
at a field-independent temperature, because the shoulders appear at the same
position in different fields. The field-dependence of magnetization for
[GdFe] was measured at $1.5$K, and a small but clear hysteresis loop was
observed as shown in the inset of Fig. 2(a). This is an intrinsic behavior,
not induced by impurity, because a single-crystal sample, which was examined
by X-ray analysis to be in a single phase, was used in measurements.
Combining these facts we could identify that the system might be magnetic
ordering with Curie temperature $T_{c}$ estimated to be around $3$K. This
will be further verified by the specific heat measurement. For a comparison,
the temperature dependence of $\chi _{M}T$ for [YFe] and [GdCo] were also
measured at $1$kOe and $5$kOe, respectively, as shown in Fig. 2(b). Since Y$%
^{3+}$ and Co$^{3+}$ [11] are diamagnetic ions, the results for [YFe] reveal
the contribution from Fe-Fe interactions, while those for [GdCo] reveal the
contribution from Gd-Gd interactions. From the observed results one may
judge that the interactions between Fe$^{3+}$ (or Gd$^{3+}$) ions are AFM in
character, whereas, the increase of $\chi _{M}T$ below $4$K for [GdFe] when
the field is less and equal to $1$kOe, suggests a weak ferromagnetic (F)
interaction between Gd and Fe ions. The magnitudes of the couplings between
metal ions are overall small, being orders of several wave numbers [12-14],
which can be estimated\ to be $|J_{Gd-Fe}|\sim |J_{Fe-Fe}|\gtrsim |J_{Gd-Gd}|
$, if the localization property of f electrons is considered. Since Gd$^{3+}$
ions have large moments, the dipolar interactions between Gd-Gd, though
still small, should not be ignored apart from superexchange interactions.
However, the dipolar interactions between Fe-Fe and Gd-Fe can be ignored, as
Fe$^{3+}$ ions are in low spin states ($S=1/2$), and the separations between
metal ions are not so small. The combination of these interactions leads
eventually to the magnetic ordering observed.

The AC susceptibility of [GdFe] was first measured in zero DC bias field
down to the lowest temperature limit $1.5$K for our measurement system
(Oxford MagLab 2000), as shown in the upper panel of Fig. 3(a). No frequency
dependent cusp was observed for either the in-phase component $\chi
_{M}^{\prime }$ or the out-of-phase component $\chi _{M}^{\prime \prime }$.
As is seen, $\chi _{M}^{\prime \prime }$ is negligibly small, while $\chi
_{M}^{\prime }$ decreases with increasing temperature, and is independent of
frequency. Since no remarkable structural disorder was found in [LnM], the
experimental measurements rule out the possibility that the system under
interest is a spin glass. However, a shoulder was observed in $\chi
_{M}^{\prime }$ versus $T$ around $3$K which is frequency independent, being
almost the same temperature at which a shoulder appears in DC measurements
in $\chi _{M}T$ versus $T$ presented in Fig.2(a), which shows that this
singularity is unique and intrinsic. This particular temperature is nothing
but the Curie temperature $T_{c}$. Then, can the nonzero applied DC magnetic
fields influence the dynamic behaviors of the magnetic compound, especially
when the field is stronger than $5$kOe? The answer is replied in the lower
panel of Fig. 3(a). When the intermediate DC bias field ($>5$kOe) was
applied, a very interesting phenomenon occurs: $\chi _{M}^{\prime }$ first
decreases to a minimum roughly at temperature $T_{c}$, and then increases to
a maximum at a finite temperature, denoted by $T_{p}$, and then decreases
with increasing temperature, while $\chi _{M}^{\prime \prime }$ first
increases, reaching a maximum, and then decreases to vanishing with
increasing temperature. When the frequency is increased from $133$ to $9333$%
Hz, the basic shapes of $\chi _{M}^{\prime }$ versus $T$ remain unaltered,
but the magnitudes become about $40\%$ smaller and the positions of peaks
move to higher temperatures in the range of $5-11$K. Apart from that the
positions of peaks, similar to $\chi _{M}^{\prime }$, also move to higher
temperatures, the magnitude of $\chi _{M}^{\prime \prime }$ exhibits
different frequency dependent behaviors, i.e., it becomes larger with
increasing frequency. From Fig. 3(a) it can be presumed that there might
exist two distinct transitions, one occurring at $T_{c}$ which is field and
frequency independent, and another occurring at $T_{p}$ which depends
strongly on field and frequency. The former may indicate an occurrence of
magnetic long-range orderings at $T_{c}$, while the latter may indicate the
presence of an unknown transition which is closely tied to the observed
unusual magnetic relaxation.

The frequency dependence of AC susceptibilities $\chi _{M}^{\prime }$ and $%
\chi _{M}^{\prime \prime }$ for [YFe] were also measured for a comparison in
zero and $1$kOe field respectively, as shown in Fig. 3(b). Zero-field AC
susceptibility for [YFe] is very similar to that for [GdFe], i.e., $\chi
_{M}^{\prime \prime }$ was detected to be almost zero, and no frequency
dependence of $\chi _{M}^{\prime }$ and $\chi _{M}^{\prime \prime }$ was
observed. As regards the AC susceptibility in a DC bias field for [YFe], the
measured $\chi _{M}^{\prime }$ and $\chi _{M}^{\prime \prime }$ are shown in
the lower panel of Fig.3(b). It can be seen that both $\chi _{M}^{\prime }$
and $\chi _{M}^{\prime \prime }$ increase from vanishingly small value to a
maximum and then decrease with increasing temperature. However, only one
peak was observed for $\chi _{M}^{\prime }$ versus $T$ in [YFe], but in
[GdFe] there are first a minimum and then a peak observed. Moreover, $\chi
_{M}^{\prime }$ approaches to the same value when $T<3$K for different
frequencies, implying that the AC susceptibility for [GdFe] has no
frequency-dependence at $T\leq 3$K. Similar measurements were carried out
for [GdCo] under zero and $5$kOe DC fields, respectively, as shown in Fig.
3(c). Apparently, [GdCo] shows even stronger field and frequency
dependences. This seems to suggest that the weak interactions between
long-distanced metal ions (Gd-Gd/Fe-Fe: $9.75-10.89$\r{A}) may play a
crucial role in such an unusual magnetic relaxation in [GdCo] and [YFe],
respectively. 

To confirm the phase transition really existing in [GdFe], the heat capacity
of pressed microcrystalline sample was measured in different DC fields using
a MagLabHC microcalorimeter (Oxford Instruments, UK). Fig. 4 presents the
temperature dependence of the total heat capacity ($C$) including the
contribution from the lattice. From the inset it can be seen that in zero
field, an anomaly was clearly observed at ca. $2.6$K. The position of this
anomaly is surprisingly consistent with the positions of shoulders observed
in DC and AC susceptibility measurements respectively, probably indicating
an onset of spontaneous magnetic long-range ordering at $T_{c}$ which is
independent of frequency and field. When further cooling below $2$K, $C$
increases very fast, and no maximum is reached down to $0.5$K which is the
working limit for our calorimeter. This singularity was not seen in DC and
AC susceptibility measurements down to $1.5$K. However, on the basis of
analyses in physics, there should be a maximum appearing in $C$ vs. $T$
curve in zero field below $0.5$K, also in order to consist with the results
of $C$ obtained in nonzero applied fields. This maximum should occur at $%
T_{p}$, marking the onset of the unknown transition caused by contributions
of frustration of metal ions. When a DC field is increased, the peaks of the
specific heat appear, and move to higher temperatures from $1$K ($10$kOe) to 
$1.5$K ($20$kOe), then ca. $2$K ($40$kOe), and finally disappear in a high
field ($80$kOe), indicating that the high magnetic fields smear the peaks
out. By combining the heat capacity and DC, AC data, we may say that this
transition is strongly field and frequency dependent, and can be probably
coined as ``magnetic relaxation phase'' because it is unusual compared with
spin glass (SG) and surperparamagnet (SP).

In addition, the novel magnetic relaxation may be understood from other
aspects. In zero bias field, the frequency dependent peaks of $\chi _{ac}$
exist in SG as well as in SP, but are absent in [LnM]. In a word, $T_{p}$ in
[LnM] occurs only after a magnetic field was applied, which is obviously
different from the behaviors in SG and SP. Although the peaks of $\chi _{ac}$
for SG and SP also show frequency dependence, but the frequency dependence
of $\chi _{M}$ in our [LnM] compounds is rather strong and slow. If we
calculate the value of relative variation of peak temperature ($T_{p}$) per
decade of frequency, $\phi $ $=\Delta T_{p}/(T_{p}\Delta (logf))$, $\phi $
is $0.53$, $0.43,1.22$ for [GdFe], [YFe] and [GdCo] in DC fields $5$kOe, $1$%
kOe and $5$kOe, respectively, while the typical value for SG is normally
less than $0.1$. If we invoke Ne\`{e}l's model which is normally assumed for
isolated SP particles to estimate the magnetic relaxation time $\tau _{0}$,
we find $\tau _{0}$ $=2.1\times 10^{-7}$s for [GdFe] in $5$kOe field; $%
9.2\times 10^{-7}$s for [YFe] in $1$kOe field; $2.7\times 10^{-5}$s for
[GdCo] in $5$kOe field. The value of $\tau _{0}$ is ca. $4-6$ orders larger
than that obtained for normal SP particles [15], showing rather slow
magnetic relaxation. These facts imply that the magnetic relaxation in these
complexes is really unusual, neither SG nor SP behaviors. Another fact must
be mentioned that the fitting results for the peaks of $\chi _{M}^{\prime
\prime }$ also give an ``energy barrier'' $E/k_{B}$, namely, they are $36$K
for [GeFe] in $5$kOe field, $27$K for [YFe] in $1$kOe field, and $24$K for
[GeCo] in $5$kOe field. It is very interesting that $\phi $, $\tau _{0}$ and 
$E/k_{B}$ for these [LnM] compounds are comparable with those in
single-molecule magnet, such as Mn$_{12}$-Ac whose $\phi $, $\tau _{0}$ and $%
E/k_{B}$ are $0.23$, $2.1\times 10^{-7}$s, and $64$K, respectively [16].
However, the present magnetic systems have extended structures than an
isolated molecule.

Here, we would like to point out that this field-dependent unusual magnetic
relaxation seems to be a rather general phenomenon in many magnetic
molecular systems with extended structures, not only limited to the systems
presently studied . We have investigated several other systems, including
the quasi-dimer [GdMnDTPA], 1D [LnMnDTPA] [17], 2D [Ln$_{2}$M$_{3}$EDTA]
[18], [LnCu] [19], [NdCo] [20], and 3D [Nd$_{2}$Co(EGTA)] [20], etc. They
all show behaviors similar to those reported above. Why do they reveal so
similar behaviors? From the structural point of view, the connection of
metal ions in these compounds shares a common character, namely,
geometrically triangular arrangements for metal ions in a layer, forming
frustrated structures. By considering this, we can infer that the physical
source for this unusual magnetic relaxation observed in the new complexes
may result from frustration which suppresses the long-range ordering
generating correlated spin clusters with slow fluctuations. However, the key
questions remain: why do the systems not exhibit any magnetic relaxation in
absence of an applied field? How does the intermediate field induce the
magnetic relaxation? We could offer a brief yet tentative argument for the
two questions. When the applied magnetic field is zero or small, the
magnetic interactions between the metal ions are overall rather weak due to
the weak Ln-M, M-M and Ln-Ln interactions (owing to the localization of f
orbitals and large M-M separation), concealing the frustration unobserved.
When the magnetic field is increasing, the AFM short-range correlations
between metal ions are increasing and dominant, as experimentally revealed,
which gives rise to frustrations unconcealed due to the coupled triangular
arrangements between AFM transition metal or lanthanide ions. In other
words, this kind of frustration is somehow unveiled by the magnetic field,
as partial degeneracies of the system can be lifted by an intermediate
field, and the phenomenon of the observed unusual slow magnetic relaxation
disappears when the applied field is strong enough. On account of this, the
frustration may be an important ingredient responsible for the observed
unusual magnetic relaxation in the compounds. The frustrated system
possesses highly degenerate ground states which are separated by energy
barriers with order of a few kelvin, and external magnetic fields can
destroy the ground state, causing the spin glass-like behavior observed. Our
findings might be a universal phenomenon existing in the weak-interacting
magnetic systems so long as frustration is geometrically present.
Furthermore, the two distinct transitions are presumed from the DC, AC
susceptibility measurements and the heat capacity data in [GdFe]. One
transition is the usual order-disorder phase transition at Curie temperature 
$T_{c}$ which is field and frequency independent, and could be attributed to
the contributions of Gd-Fe interactions since no long-range ordering was
observed in [YFe] and [GdCo]; while another transition occurring at
temperature $T_{p}$ which depends strongly on field and frequency, is
unknown at the moment, but we could speculate that it might be closely
related to the unusual magnetic relaxation. Seemingly, these two transitions
have distinct mechanisms and no relation. Since the results uncovered in
these compounds are quite complicated, a simple theoretical model is not
feasible now. But we hope that a proper theory could be sooner established
to explain these unusual behaviors.

\acknowledgements

We are supported in part by the National Natural Science Foundation of
China, and State Key Project for Fundamental Research. We are grateful to
Prof. Dr. H. Lueken for some measurements on susceptibility. S.G. and G.S.
acknowledge supports from the Alexander von Humboldt Stiftung.

\bigskip

FIGURE CAPTIONS\newline

Fig.1. (a) Illustration of Fe-CN-Gd chains and connections of Fe$^{3+}$ ions
(solid line) and Gd$^{3+}$ ions (dash line) in {\it ab} layer for [GdFe].
(b) The connection of Fe (or Gd) ions in and between layers.

Fig.2. (a) $\chi _{M}T$ versus temperature ($T$) for [GdFe] in different
fields. Inset: hysteresis loop for [GdFe] at $1.5$K. (b) $\chi _{M}T$ versus
temperature ($T$) for [GdCo] and [YFe] in different fields.

Fig.3. Temperature dependencies of the in-phase AC magnetic susceptibility, $%
\chi _{M}^{\prime }$, and the out-of-phase AC susceptibility, $\chi
_{M}^{\prime \prime }$, in absence (upper) and presence (lower) of a bias DC
field for different frequencies. (a): [GdFe]; (b): [YFe]; (c): [GdCo].

Fig.4. Temperature dependence of the heat capacity for [GdFe] in different
fields. Left inset: Enlarged plot in temperature range of $0-6K$; Right
inset: Enlarged plot in temperature range of $2-3.0K$.


\begin{references}
\bibitem{11}  Note that the octahedral CN surroundings induce a low-spin
state of d electrons localized on transition-element cation. The six d
electrons of Co$^{3+}$ occupy T$_{2g}$ orbitals and are thus paired, leading
to Co$^{3+}$ ions exhibiting diamagnetic.
\end{references}
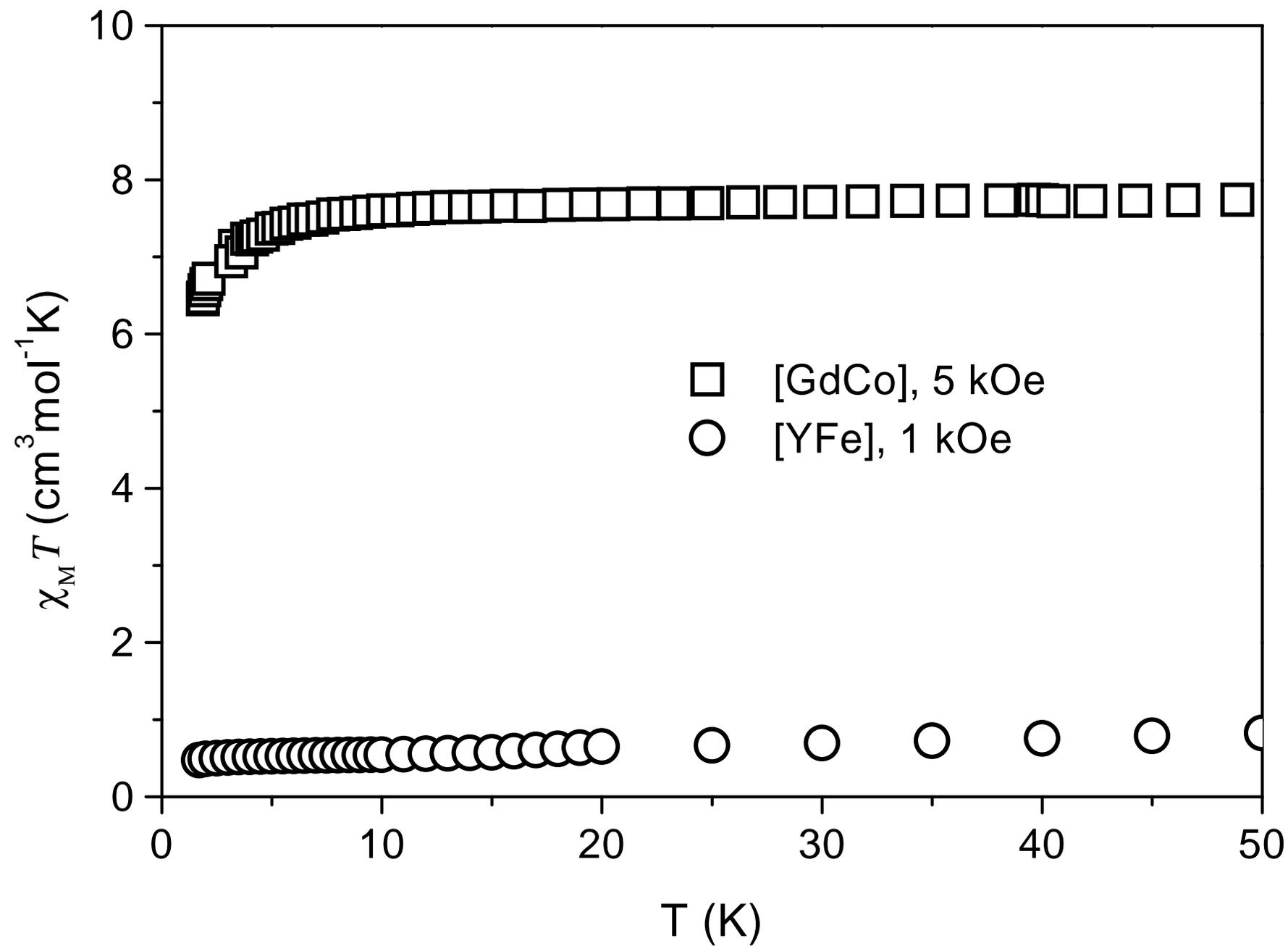
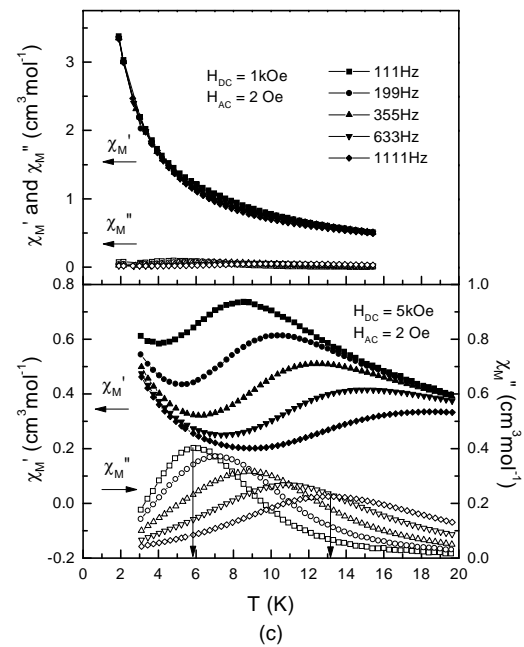
\end{document}